# ДРОПШИППИНГ – АЛЬТЕРНАТИВНАЯ ИНФРАСТРУКТУРА СБЫТА И ПРОДВИЖЕНИЯ

Калужский М.Л.
к. ф. н., доцент ОмГТУ, e-mail: frsr@inbox.ru

***Аннотация***: В статье анализируются социально-экономические последствия развития дропшиппинга – новой формы организации продаж, стремительно завоевывающей российский рынок. Дропшиппинг открывает огромные перспективы не только для простых людей, но и для российских производителей, которые при правильном использовании возможностей дропшиппинга могут с минимальными затратами получить прямой доступ к мировому рынку.

***Ключевые слова***: интернет-маркетинг, дропшиппинг, экономический кризис, глобализация, интернет-аукцион, электронная торговля площадка, платежные системы.

# DROPSHIPPING - ALTERNATIVE INFRASTRUCTURE OF SALES AND PROMOTION
*Kaluzhsky M.L.*

***Abstract***: *An article about the transformation of the theory and practice of marketing in terms of e-commerce and network economy. The author considers Internet Marketing as an independent marketing communication in a virtual environment. The main thesis of the article: virtual environment determines the transformation of marketing, changing methods, priorities and structure not only practice, but also the theory of marketing.*

***Keywords***: *internet marketing, dropshipping, the economic crisis, globalization, online auction, e-commerce platform, payment systems.*

**Докризисный маркетинг.** Начало 2000-х гг. ознаменовалось бурным развитием транснациональных интеграционных процессов в экономике и глобализацией международных рынков. Ситуация, когда товар разрабатывался в США, производился в Китае, а продвигался на рынок в Европе, стала повсеместной нормой. При этом укрупнение производства и развитие международного разделения труда привели к постепенному отмиранию или поглощению крупными компаниями мелких и средних производств. Только крупные холдинги и транснациональные компании могли успешно конкурировать в условиях глобального рынка.

Одновременно н докризисный период происходили ускоренное становление крупных розничных торговых сетей, быстрое обновление товарного ассортимента и неуклонный рост совокупного потребления. Глобализация стола признаком не только массового производства товаров, но и их продвижения [1]. По мере укрупнения производств происходило укрупнение рыночной инфраструктуры, глобализация рынков и маркетинговых кампаний по продвижению продукции.

В маркетинговой практике на этот период приходятся бурный расцвет крупных торговых сетей, создание массовых производств в Китае к странах Юго-Восточной Азии. Страны Юго-Восточной Азии и Китай превратились в гигантскую фабрику. Этот период характеризуется повальным увлечением новыми формами продвижения и позиционирования товаров; брэндингом, мерчандайзингом, франчайзингом, факторингом и т.д.

Традиционный маркетинг постепенно утрачивал свое значение, выйдя за рамки традиционной конкуренции. В самом деле, зачем тратить силы на использование методов традиционного маркетинга, если простой перенос производства из США в Китай давал от 300 до 1.500% чистой прибыли за счет экономии на стоимости рабо-чей силы. При такой рентабельности можно было просто «завалить» потребителей своими товарами.

Бурный рост потребления обеспечивался за счет массового кредитования Федеральном резервной системой США мирового производства и потребления на всех стадиях производственно-потребительского цикла. Все это привело к повсеместной деградации традиционных форм торговли, сращиванию финансовых и торговых структур, бурному развитию крупных холдингов и вертикально интегрированных компаний.

Российская экономическая практика тут не была исключением из общемировых тенденций. Схема была довольно простая:

1. Один из участников цепи товародвижения подучал существенное преимущество по отношению к своим партнерам в виде монопольного положения в сбытовой сети.

2. Максимизируя свою прибыль и минимизируя издержки, он постепенно приобретал контроль нал своими контрагентами.

3. В итоге формировалась вертикально интегрированная компания, включающая всех (или почти всех) участников производственно-торговой цепочки на пути товара от производителя к потребителю.

Это общемировая тенденция докризисного периода. Огромным минусом такого развития событий стали монополизация торговли и массового производства крупными холдингами, удушение свободной конкуренции и вытеснение с рынка мелких и средних участников. Здесь тоже был свой пузырь, который лопнул в 2008 г.

Основными факторами успешности на этом этапе была не маркетинговая политика, а доступ субъектов к источникам кредитных ресурсов н преимущества укрупнения бизнеса. Однако доступность кредитных ресурсов не только стимулировала номинальный экономический рост, но и вела к серьезнейшим дисбалансам между доходами и расходами, как домохозяйств, так и самих хозяйствующих субъектов, Практически все жили в долг. Достаточно сказать, что основным источником капитализации большинства транснациональных высокотехнологичных компаний в докризисный период стала эмиссия ценных бумаг, а вовсе не их производственный потенциал. Потом жизненный цикл сложившейся системы отношений завершился, и она рухнула.

**Мировой экономический кризис**. С переходом мирового экономического кризиса в острую фазу ситуация на потребительских рынках кардинально изменилась:

L Разрушение системы потребительского кредитования в разы уменьшило покупательную способность потребителей и соответственно емкость потребительских рынков.

2. Крупные западные производители утратили контроль над производственными мощностями в Китае, которые перешли на самостоятельное производство товаров по западным технологиям.

3. «Брендовые товары» стали менее доступны рядовым покупателям, а на первое место в качестве стимула для принятия решения о покупке вышли цена и функциональность продукта, а не его престижность, как ранее.

4. Торговые сети столкнулись с затовариванием и замедлением оборота. Труднее всего пришлось сетям в сфере бытовой электроники. Будучи не в состоянии рассчитаться по текущим кредитам, они не могли обеспечить своевременное обновление ассортимента.

5. Докризисные методы продвижения продукции утратили свою эффективность. Ушли в прошлое масштабные PR-кампании, крупные торговые сети. Вертикально интегрированные компании разоряются, а покупатели ориентируются, прежде всего, на свои насущные потребности и реальные возможности.

В итоге кризис привел к серьезнейшим изменениям в структуре потребления, производства и продвижения товаров. Производители в попытках стимулировать спрос вынуждены ускоренными темпами выводить на рынок технологические новинки, но это лишь усугубляло положение и без того затоварившейся торговли.

В результате в мировой торговле наблюдается быстрое падение потребительского спроса, и он, по всей вероятности, будет падать в течение ближайших нескольких лет, пока экономический спад не достигнет своего дна. У этого явления совершенно объективные причины, и повлиять на них уже никому не под силу. Процесс деградации глобальной системы мировой экономики, завязанной на стимулирование потребительского спроса из единого эмиссионного центра в США, необратим.

Вместе с тем, экономический кризис несет в себе определенную революционную составляющую в маркетинговой деятельности на потребительском рынке. Появляются новые возможности продвижения продукции, меняются мотивы потребительского поведения и структура конечного спроса. Перестают работать старые схемы продвижения и позиционирования товаров на рынке, но на их месте уже сейчас возникают новые, более простые и эффективные.

Тут есть как плюсы, так и минусы. Мы стоим на пороге не просто смены экономической модели. Мы стоим на пороге смены самой сущности экономических отношений.

С одной стороны, постоянное удорожание сырья как средства вложения свободных денежных средств ведет к формированию финансовых пузырей и сдерживает появление новых производств. Себестоимость производства растет, а спрос падает. Торговые цепочки сокращаются, оптовая торговля и крупные торговые сети теряют объемы. Это ведет к разорению производств и деградации связанных с ними систем распространения.

С другой стороны, в условиях паления совокупного спроса сырьевые пузыри рано или поздно лопнут, и сырье подешевеет. Убытки понесут те компании, которые сохраняют деньги в сырьевых запасах. Они не будут поспевать за падением цен на сырье, как ОАО «Газпром» сейчас не поспевает за спотовыми ценами на сжиженный газ. Зато выиграют компании, ориентированные на реальный спрос и реальные запросы потребителей.

Развитие структурного экономического кризиса уже сегодня привело к перетоку значительной части покупателей на новые рынки с новыми формами торговли и продвижения продукции. Прежде всего, речь идет об интернет-торговле. До кризиса интернет-торговля была представлена в основном торговыми интернет-площадками для участников с небольшими объемами продаж и глобальными торговыми площадками (*Amazon*, *eBay*, *Delcampe* и др.). Кризис изменил структуру интернет-торговли в не меньшей мере, чем структуру традиционной торговли.

Он привел к деградации непомерно развившейся торговой инфраструктуры. Но одновременно бурное развитие получили электронные торговые площадки, системы электронных платежей, дропшиппинг и почти забытая посылочная торговля. Благодаря сети Интернет потребительские товары не стали менее доступны, чем ранее. Наоборот, доступный ассортимент для рядового потребителя увеличился в сотни и тысячи раз.

Благодаря сети Интернет в короткие сроки сформировались новые формы торговли с короткими каналами сбыта, минимальными оборотными издержками и минимизированным налогообложением. Именно сюда потянулись вытесненные торговыми сетями и крупными супермаркетами мелкие предприниматели. И именно здесь сегодня царят настоящий рынок и свободная конкуренция.

**Развитие дропшиппинга.** Дропшиппинг как самостоятельная форма продвижения товаров на рынке зародился задолго до кризиса в США и в первоначальном варианте подразумевал мелкооптовую торговлю через Интернет товарами массового спроса, ориентированную на мелких торговцев по всему миру. Типичным примером такой формы торговли может служить несколько архаичный интернет-портал *Buywholesalelots* (*www.buywholesalelots.com*), специализирующийся на мелкооптовой торговле товарами из США и Канады.

Грянувший мировой экономический кризис 2008 г. вызвал бурное развитие дропшиппинга. Именно дропшиппинг окончательно добил розничные торговые сети бытовой электроники, предоставив покупателям возможность приобретения технологических новинок, не выходя из дома, по цене в 1,5-2 раза дешевле магазинных.

В США и Европе возникли компании-посредники в сфере дропшиппинга, специализирующиеся на организации взаимодействия между поставщиками и дропшипперами. Например, американская компания *Doba* (*hltp://www.doba.com*) предлагает для продажи всем зарегистрировавшимся на сайте ассортимент более чем из 1.500.000 товаров от более чем 300 производителей в 8.000 категорий, включая 3.000 брендов и 7 дней бесплатного доступа к базе данных производителей товаров. В российских компаниях, например *Drop-shipping.ru* (*http://www.drop- shipping.ru*), принцип работы аналогичен.

Однако западные посредники не могут даже близко конкурировать с китайскими компаниями, у которых цены гораздо ниже, а доставка зачастую бесплатная. Поэтому основные поставщики товаров для дропшиппинга находятся сегодня в Китае, а не в Европе или США,

Четыре основных фактора, спровоцировавших возникновение этого явления на потребительском рынке:

1. Резкое паление потребительских доходов и уровня жизни населения, заставившее покупателей переориентироваться на интернет-покупки.

2. Широкая доступность интернет-технологий даже для небольших компаний, позволившая поставщикам отказаться от услуг традиционной оптово-розничной торговли.

3. Огромная ценовая вилка между отпускными и розничными ценами, поддерживаемая протекционистской политикой властей КНР по продвижению своих товаров на внутренние и внешние рынки,

4. Наличие дешевых торговых площадок. делающих интернет-торговлю доступной для любого желающего даже при отсутствии первоначального капитала.

Сегодня практически все китайские производители мелкой бытовой техники, одежды и прочих товаров массового спроса имеют на своих сайтах страницу регистрации и службу поддержки для работы с дропшипперами. Кроме того, в КНР действуют десятки (если не сотни) сайтов, не только аккумулирующих по американскому образцу информацию об осуществляющих розничную почтовую отгрузку производителях, но и самостоятельно выполняющих за них эту функцию.

Наиболее типичным в этом смысле является компания *LighTake* (*http://www.lightake.com*), которая не только аккумулирует предложения поставщиков, но и выполняет функции обработки заказов и отгрузки продукции. Своего рода интернет-магазин для дропшипперов. Причем в качестве обратного адреса на посылке указывается любой адрес по желанию дропшипперов, для которых предусмотрена особая система льгот и зачетов.

Китай в этом смысле значительно опередил своих западных конкурентов благодаря мировому экономическому кризису, Китайское правительство в качестве меры по стимулированию внутреннего потребительского спроса обязало китайских производителей продавать до 30% экспортной продукции на внутреннем рынке по себестоимости.

И без того дешевые китайские товары получили гигантскую ценовую фору при проникновении на зарубежные рынки. В кратчайшие сроки сформировалась огромная инфраструктура мелких компаний, закупающих товары оптом по бросовым ценам и перепродающих их через Интернет.

Первоначально это был спам, рассылаемый почтовыми автоматами по миллионам адресов интернет-пользователей (например, *http://www.viapaypal.com*). Эффективность такого продвижения была сравнительно невелика. Однако затем возникли мелкооптовые интернет-сайты по образцу сайта *Buywholesalelots* (*http://www.buywholesalelots.com*). Такие, например, как *Alibaba.com* (*http://www.alibaba.com*) и *DHGate.com* (*http://www.dhgate.com*). Эти торговые площадки предназначались для производителей и оптовых посредников, торгующих партиями от 10 до 1000 экз.

Параллельно уже существовали крупные торговые площадки *Amazon.com* (*http://www.amazon.com*), *eBay* (*http://www.ebay.com*) и многие другие, ориентированные на вторичный рынок. Именно там традиционно концентрировался потребительски» спрос. Однако там развитие дропшиппинга сдерживалось высокой (около 20%) суммой сборов с продаж.

Зато на более дешевых площадках (*Aukro*, *Molotok* и др.) дропшиппинг получил бурное развитие почти одновременное началом кризиса. Этому способствовал целый ряд объективных условии, и прежде всего низкий размер торговых сборов. Например, на *Молотке* (*http://molotok.ru*) он составляет 3% и менее от суммы продаж с бесплатным выставлением товара на пролажу. Это почти в 8 раз дешевле, чем на *eBay* и на *Amazon*.

В России дропшиппинг стал логическим продолжением нашего российского челночества – повального увлечения начала 1990-х гг. Там все начиналось с частных поездок в Китай российских граждан на свой страх и риск. А закончилось созданием в КНР на границе с Россией мощных торговых центров с полным комплексом услуг (включая проживание, централизованную отгрузку товара и его транспортировку до места назначения).

Благодаря дропшиппингу отпадает даже сама необходимость поездок в Китай за товаром. Достаточно зарегистрироваться на интернет-сайте интересующей компании, заполнить анкету, и весь ассортимент предлагаемой продукции немедленно поступает в распоряжение торговца.

В результате дропшиппинг развивается по трем направлениям:
1. Тематические группы в социальных сетях (*ВКонтакте*, *Facebook* и др.), с помощью которых энтузиасты рекламировали и продвигали товары с мировых интернет-аукционов.
2. Интернет-магазины, отказавшиеся от оптовых закупок из-за заговаривания и падения потребительского спроса.
3. Энтузиасты, имеющие опыт продажи подержанных товаров на интернет-площадке *Молоток* (*http://molotok.ru*), самостоятельно освоившие принципы дропшиппинга.

И если в начале кризиса предложения дропшипперов были достаточно примитивны, то сегодня это современные раскрученные интернет-магазины с высококачественным дизайном, возможностью оплаты через платежные системы *VISA*, *MasterCard*, *WebMoney* и др. Это кардинально меняет всю систему и инфраструктуру современной торговли. Расстояние до производителя теперь не имеет решающего значения. Страновые границы стираются, а любой, даже самый мелкий производитель, имеет возможность свободно выходить на зарубежные рынки, не покидая своего офиса.

**Сущность дропшиппинга**. Дропшиппинг (от англ. dropshipping) – прямая поставка, своего рола «интернет-консигнация». Посредник (дропшиппер) продает то-вары поставщика от своего имени, оформляя заказ на поставку после получения оплаты от покупателей. Затем деньги переводятся поставщику, который сам отгружает товар клиенту.

Получается схема, с успехом применяемая производителями мороженого, когда товар передается на консигнацию, киоск сдается и аренду, а киоскер является формально независимым частным предпринимателем. Здесь эта схема доведена до абсолютного совершенства, поскольку поставщик без каких-либо внешних вложений получает огромное количество работающих на свой страх и риск торговых представителей. В общих чертах схема работы дропшиппинга выглядит следующим образом:

*Этап 1*. Посредник находит сайт поставщика товаров, цена на которые в разы отличается от цен на местных рынках при условии, что поставщик продает свои товары почтой.

*Этап 2*. Посредник делает пробную закупку и обговаривает с поставщиком условия сотрудничества (гарантия, условия отгрузки и т.д.).

*Этап 3*. Посредник копирует описание и изображения товаров на сайте поставщика или делает их самостоятельно, а затем выставляет товары продавца на электронных торговых площадках от своего имени.

*Этап 4*. Покупатели приобретают товары у посредника с отгрузкой от поставщика. Причем посредник только принимает заказы и переводит оплату за минусом свой комиссии поставщику.

*Этап 5*. Поставщик отгружает оплаченные товары по адресам, предоставленным ему посредником, и с ним же решает вопросы, связанные с гарантией на проданные товары.

С точки зрения маркетинга такая схема позволяет поставщику без особых затрат быстро выйти на любые рынки независимо от страновых или иных различий. Оптово-розничная торговля больше становится не нужна. Большой штат торговых работников, промежуточные склады, логистика поставок, сложные договорные отношения также не требуются.

Если появляется товар, объективно востребованный на потребительском рынке, то он мгновенно распространяется по рынку. Функции отдела сбыта продукции трансформируются в функции администрирования интернет-сайта, когда основная задача менеджмента заключается в организации взаимодействия с дропшипперами и своевременной обработке поступающих от них заявок.

В самом деле, зачем нужны розничные магазины, оптовые базы и сопутствующий торговый персонал, если товар продается напрямую покупателям? В этой системе торговли не бывает неплатежей контрагентов, проблем с распределением товара, неэффективных транспортных расходов и прочих «прелестей» традиционной торговли.

Данный вид бизнеса практически не требует от дропшипперов начального капитала. Следовательно, дропшиппер не несет практически никаких предпринимательских рисков. Оплату товара поставщику он осуществляет лишь после того, как получает предоплату от покупателя. Среди других преимуществ дропшиппинга для посредников можно выделить следующее:

1. ***Дропшипперу не требуется складских помещений***. Все, что нужно ему для организации продаж, – персональный компьютер с выходом в сеть Интернет и наличие договоренности с поставщиком,

2. ***Все заботы по отправке товара берет на себя поставщик***. Он же предоставляет номер для отслеживания почтового отправления, осуществляет гарантийный ремонт, замену продукции и предоставляет необходимую информацию о товаре.

3. *Посредник имеет возможность сотрудничать с любым количеством поставщиков* одновременно на любом количестве торговых площадок, расширяя ассортимент товаров и наращивая количество клиентов.

4. *Поставщик отправляет товары покупателям от имени дропшиппера*, который имеет возможность создавать узнаваемую торговую марку и самостоятельно продвигать ее в сети Интернет,

Однако в дропшиппинге нет и не может быть обязательств торгового представителя перед поставщиком. Они вступают в экономические отношения в момент перевода денежных средств и поступления заказа от посредника поставщику. Поэтому дропшиппинг доступен всем желающим, обладающим минимальными навыками интернет-торговли. Здесь царит свободный рынок, а предпринимательский успех определяется уровнем маркетинговой подготовки дропшипперов.

**Маркетинг в дропшиппинге**. В условиях свободной конкуренции и равного доступа дропшипперов как к поставщикам, так и к потребителям, основным фактором конкуренции становится уровень индивидуальной маркетинговой подготовки дропшиппера, его практические знания и навыки. Здесь складывается несколько иная, отличная от традиционной, технология маркетингового продвижения товаров.

*Во-первых*, товар как первый элемент традиционного комплекса маркетинга учитывается, что называется, «при прочих равных». С одной стороны, все конкуренты дропшиппера имеют равный доступ к товарам поставщиков. С другой стороны, потребители через сеть Интернет обладают равными возможностями по получению информации о товаре, его свойствах и конкурентных преимуществах (в том числе на тематических форумах).

Это ведет к тому, что для дропшиппера использование данного элемента комплекса маркетинга как инструмента продвижения ограничивается его личными возможностями. Он, конечно, может увеличить срок гарантии или дополнить товар дополнительными комплектующими (например, картами памяти или батарейками). Однако делать этого он будет на свой страх и риск и за свой счет. Поставщик никаких дополнительных возможностей в сфере товарной конкуренции дропшипперу не предоставляет.

Таким образом, свойства товара в дропшиппинге – это инструмент продвижения поставщика (производителя), но ни как не дропшипперов. Рынок очень быстро сведет на нет все преимущества посредника, сделавшего ставку на уникальность своего торгового предложения.

*Во-вторых*, цена как второй элемент комплекса маркетинга является важнейшим фактором продвижения товара. Причем поскольку дропшипперы находятся изначально в равных условиях, то и цена является в первую очередь инструментом продвижения поставщика.

Задача поставщика – обеспечить такую разницу между справедливой ценой в сознании потребителей и своей отпускной ценой, чтобы предлагаемый товар стал привлекателен для потенциальных дропшипперов. Далее вступает в силу закон эластичности спроса по цене, и объемы закупок (продаж) будут регулироваться лишь потребительским спросом на рынке.

В Китае, где функции производства и сбыта зачастую разделены между контрагентами, существует система сбыта, в рамках которой инвесторы выкупают большие партии высоколиквидных товаров у производителей, а затем перепродают их дропшипперам и конечным потребителям от своего имени. Иначе говоря, если «ценовая вилка» между отпускной и розничной ценой достаточно велика, то сбытовая инфраструктура выстраивается сама собой. Это – одно из важных преимуществ глобального рынка и интернет-торговли.

***В-третьих***, распределение как третий элемент комплекса маркетинга является неотъемлемой прерогативой дропшипперов. Возможности поставщика здесь крайне ограничены и сводятся к построению единой логистической схемы приема и обработки заказов.

Вместе с тем, именно дропшипперы выступают в качестве активного элемента системы товародвижения в дропшиппинге. Они на свой страх и риск осваивают новые целевые рынки, торговые площадки, платежные системы и т.д. Интернет-торговля ограничивается лишь доступом аудитории к сети, и потому даже самый мелкий торговец при удачном стечении обстоятельств может показать выдающиеся результаты.

В сферу сбытовой конкуренции здесь попадают, прежде всего, методы доведения информации до целевой аудитории. Причем, речь идет не о рекламе. Основная часть покупателей ищет конкретный товар, и выигрывает тот продавец, предложение которого окажется в нужном месте в нужное время. Отсюда главная задача сбытовой политики в дропшиппинге заключается в доведении предложения о продаже товара до максимального количества потенциальных потребителей при помощи сети Интернет.

***В-четвертых***, продвижение как четвертый элемент комплекса маркетинга, так же как и первые два элемента, является прерогативой в первую очередь поставщика (производителя) товара. Потенциал дропшипперов здесь крайне ограничен и силу их абсолютно равных маркетинговых возможностей.

Зато возможности производителей эксклюзивных, не имеющих аналогов товаров почти безграничны. Начать можно с того, что стоимость интернет-рекламы гораздо ниже, чем стоимость общенациональных рекламных кампаний, тогда как аудитория её весьма и весьма велика.

Это значит, что производитель может попытаться за существенно меньшие деньги заранее сформировать у интернет-аудитории завышенное представление о предлагаемом товаре и его цене (например, создать сверхпосещаемый специализированный сайт). Затем товар можно предлагать дропшипперам для продажи, и они уже сами методами прямых продаж (сетевого интернет-маркетинга) доведут информацию о «достоинствах» товара до потенциальных потребителей. В России такая схема уже реализована, к примеру, на сайте «*Система продаж «Продамус»*» (*http://www.prodamus.ru*).

Маркетинговые исследования в дропшиппинге проводятся на основных торговых площадках и сводятся в основном к анализу потребительского спроса (это несложно отследить по рейтингу продавцов), анализу конкуренции и ценовых категорий (через анализ предложений). Огромным плюсом дропшиппинга является то, что благодаря коротким каналам сбыта поставщик может отслеживать спрос на рынке, что называется «в реальном времени».

С позиций дропшипперов исследования сводятся к поиску поставщиков высокорентабельных товаров и выявлению неохваченных ниш на потребительском рынке. С позиций поставщиков исследования сводятся к поиску рыночных возможностей по созданию привлекательных условий для дропшипперов.

В любом случае доминирует здесь не маркетинг, воздействующий на потребителей (как было до кризиса), а маркетинговая адаптация к уже существующим потребностям. Сначала производитель (поставщик) выявляет потребительский спрос, затем он позиционирует товар, ориентированный на удовлетворение этого спроса, и только потом дропшипперам распространяют товар на потребительском рынке.

**Инфраструктура дропшиппинга**. Основным фактором успешного бизнеса в дропшиппинге является наличие развитой инфраструктуры сбыта. Наибольшие проблемы поставщиков здесь связаны с тем, что в дропшиппинге дропшипперы не связаны с поставщиком договорными обязательствами. Сегодня им выгодно работать с поставщиком, но завтра все может стремительно измениться ...

Регулируется этот процесс факторами, не зависящими от поставщика, но хорошо поддающимися учету и анализу. Речь идет о технологических возможностях и сервисных функциях, реализуемых сторонними компаниями, без которых дропшиппинг невозможен.

1. *Электронные торговые площадки* – серверы, на которых покупатели и продавцы совершают электронные сделки по купле-продаже товаров через сеть Интернет. Принцип работы торговой площадки достаточно прост: купить товар может любой зарегистрировавшийся покупатель у любого зарегистрированного продавца, Отказ от сделки или невыполнение ее условий ведет к дисквалификации участника. Порядочность участников подтверждается системой взаимных отзывов. Для покупателей все сделки бесплатны.

Самой крупной в мире торговом площадкой (50 млн. пользователей) является интернет-аукцион *eBay* (*http://www.ebay.com*), объединяющий, помимо основной площадки, еще и дочерние площадки в 30 странах. Купить там можно практически все: от антиквариата и одежды до автомобилей и недвижимости.

Давно и успешно действуют на потребительском рынке также площадки *Amazon* (*http://www.amazon.com*), *Delcampe* (*http://www.delcampe.net*), Taobao (*http://taobao.com*) и множество других. Существует и российский аналог интернет-аукциона *eBay –Molotok* (*http:// www.molotok.ru*), входящий в международную группу *MIH Allegro B.V.* Эта компания поддерживает торговые площадки в 10 странах Европы н СНГ, обслуживая около 7 млн. зарегистрированных пользователей.

Действуют торговые площадки по принципу биржи, а финансируются за счет отчислений продавцов от продаж. Так, на аукционе *eBay* эти отчисления самые большие в мире: около 20-25%. Другие электронные торговые площадки более демократичны (сборы составляют от 3 до 5% стоимости проданного лота), но пока за счет меньшего количества участников они менее привлекательны.

Создание электронных торговых площадок – общая тенденция последних лет, как за рубежом, так и в России. Активно функционируют на российском рынке торговые площадки отдельных предприятий (например, *ОАО «Северсталь»*, *http:// torg-severstal.ru*), региональные площадки (*http://prodam.slando.ru* и др.), а также отечественные компании, разрабатывающие программное обеспечение для такого рода деятельности (например, *ПП «Госзакупки»*, *http://www.etc.ru*) и др.

Маркетинговая ценность электронных торговых площадок заключается в возможности быстро и с минимальными затратами донести информацию до огромной целевой аудитории. Электронные торговые площадки продают торговцам две услуги – доступ к аудитории и возможность продавать с минимальным набором навыков работы в сети Интернет. Покупателям же электронные торговые площадки предлагают такой ассортимент товаров, который не способен предложить ни один, даже самый крупный, самостоятельный продавец. В этом заключается их основное конкурентное преимущество в сравнении с традиционными интернет-магазинами.

2. *Платежные системы* – это традиционные системы банковских переводов и электронные платежные системы, позволяющие клиентам совершать мгновенные платежи, не выходя из дома (офиса). Принцип работы электронной платежной системы заключается в передаче информации о совершаемом платеже через сеть Интернет в процессинговый центр, который осуществляет расчеты в режиме реального времени. Экономится не только время, но и затраты на кассовое обслуживание клиентов, инкассацию, оборудование и т.д.

Самыми крупными в мире электронными платежными системами являются *PayPal* (*www.paypal.com*) и *Moneybookers* (*www.moneybookers.com*), обслуживающие десятки миллионов держателей счетов по всему миру. Стоимость их услуг колеблется в пределах 3% от суммы перевода. Существуют и другие платежные системы (*BidPay*, *Mi-Pay* и др.), но их аудитория сравнительно невелика. В целом можно сказать, что основным фактором успешности платежной системы начнется ее привязка к действующей электронной торговой площадке, как, например, привязка *PayPal* к *eBay* или привязка *Moneybookers* к *Delcampe*.

В России и СНГ самой распространенной платежной системой является *WebMoney* (*www.webmoney.ru*), осуществляющая операции в рублях, валюте и золоте, а также предоставляющая пользователям возможности взаимного кредитования. Стоимость услуг по переводу денег для зарегистрированных пользователей *WebMoney* составляет 0,8% от суммы платежа.

Помимо *WebMoney* на российском рынке действует множество других платежных систем, предоставляющих клиентам услуги по осуществлению электронных переводов (*Yandex-деньги*, *RBK-Money*, *Z-Payment*, *LiqPay* и др.), приему электронных платежей на сайте компании (*ASSiST*, *Active Pay*, *OnPay*, *QuickPay* и др.), приему платежей в пользу сторонних организаций (*CyberPlat*, *Новоплат*, *e-Port* и пр.).

Кроме того, на российском рынке платежи за товары можно осуществлять через:

а) традиционные платежные системы на основе банковских карт (*VISA*, *MasterCard*) и их российские аналоги (*Золотая Корона*, *Сберкарт* и др.);

б) систему почтовых переводов (*КиберДеньги* и *Форсаж*);

в) банковские переводы по системе *SWIFT*:

г) пополнение банковского счета или счета мобильного телефона;

д) суррогатные карты - например, платежные карты торговой сети «*Евросеть*» (*http://kykyryza.ru*);

е) международные платежные системы (*MoneyGram*, *Western Union*. *Contact*, *Unistrim*, *Anelik*, *Begom*, *Migom* и др.).

По статистике, чаще всего покупатели оплачивают покупки почтовыми переводами, хотя именно *ФГУП «Почта России»* берет за свои услуги наибольшую комиссию (5-7% плюс 25 руб. при сумме до 1.000 р.). На втором месте по популярности стоит система денежных переводов *Contact* с 1,5-3% от суммы без минимума. Затем следует *WebMoney*, но здесь покупатели предпочитают оплачивать через платежные терминалы компании-посредников, а это прибавляет еще 3-5% к стандартным 0,8% от суммы платежа.

Вариантов приема оплаты от покупателей на рынке присутствует великое множество. Однако многие торговцы часто забывают, что для покупателя в условиях кризиса именно цена является решающим фактором при принятии решения о покупке. Сегодня существуют вполне легальные возможности бесплатного приема денежных средств через пополнение банковских карт, предоставляемые некоторыми банками (например, «*СМП-банк*», «*ВТБ-24*» и др.). Эти методы вряд ли вытеснят с рынка традиционные формы оплаты, но повысить привлекательность коммерческого предложения вполне в состоянии.

3. **Посреднические компании** – это компании, специализирующиеся на сервисном сопровождении покупок за рубежом. Они специализируются на покупке, оформлении и доставке в Россию товаров от имени и по поручению клиентов. Крупнейшие компании в России: *Ebay Today* (США, Англия; генерирует для России 3-4 тыс. EMS-отправлений в месяц), *Shipito* (США; 8-10 тыс.), *YOOX* (Италия; 1,5-2 тыс.), *G-Market* (Южная Корея; 3,5-4.5 тыс.).[1]

Стандартные услуги таких компаний:

---

[1] По материалам Министерства связи и массовых коммуникации Российской Федерации. – WEB:http://minsvyaz.ru/ru/monitoring/index.php?id_4=42227.

- предоставление виртуального адреса для клиентов в стране пребывания продавца товара (например, *eBay Today* предоставляет адреса в США. Германии, Великобритании, Японии и Китае);
- прием, хранение, группировка и почтовая отправка покупок со склада в стране пребывания продавца товара;
- покупка товаров на интернет-аукционах и в интернет-магазинах со своего аккаунта по поручению клиента;
- таможенное оформление, сопровождение и получение товаров на территории Российской Федерации (в г. Москве).

Все это делает доступными покупки даже в тех случаях, когда продавец отказывается отправлять товар за границу.

В дропшиппинге речь идет, например, о том, что многие европейские компании устанавливают дискриминирующие цены на товары для покупателей внутри Европейского союза и за его пределами. Например, цены на продукцию ведущего европейского производителя электрических каминов ирландской компании *Dimplex* отличаются в российском представительстве компании и в английских представительствах минимум в 2,5 раза. По условиям контракта с производителем контрагенты в Европе не могут продавать продукцию за пределы Европейского союза по внутренним ценам. Однако они охотно продают товары на виртуальный адрес в Великобритании. Все остальное – доход дропшиппера.

4. *Транспортные компании* – это частные и государственные компании, оказывающие клиентам услуги по транспортировке (и иногда растаможиванию) товаров. Их деятельность обеспечивает то, что в советские времена называлось «посылочной» («каталожной») торговлей.

Основным агентом для получения товаров из-за рубежа по-прежнему остается *ФГУП «Почта России»*. Это очень неэффективная государственная структура, где присутствуют воровство, безответственность и наплевательское отношение к клиентам. Раздувание в последние годы управленческих штатов только усугубило ситуацию в этом ведомстве.

Государство «похоронило» советскую систему посылочной торговли в начале 1990-х гг., когда были отменены старые «Почтовые правила». По измененным правилам оплата за доставку и возврат отправленных наложенным платежом невыкупленных товаров взимается в двойном размере. Кроме того, почта не разрешает вскрывать отправлений с наложенным платежом для проверки товаров до их получения и не несет никакой реальной ответственности за хищения.

Вместе с тем, средний показатель вскрытых и разворованных почтовых отправлений пока редко превышает 2-3%, что позволяет продавцам списывать убытки за счет прибыли. Однако про отправку товаров наложенным платежом большинству торговцев пришлось забыть. Несомненным преимуществом обычной почты являются широкая доступность почтовых услуг и относительно низкие почтовые тарифы.

Параллельно обычной почтовой доставке товаров широкое развитие получила скоростная почта (EMS, DHL). По статистике ФГУП «Почта России» через EMS ежемесячно ввозятся в Российскую Федерацию около 100 тыс. отправлений в месяц. 30% из них приходится на США, 20% – на Китай, 10% – на Южную Корею, 1% – на Японию, 4,5% – на Израиль и 4% – на Великобританию.[2]

По скорости доставки скоростная почта недалеко ушла от обычной почты, и условия оказания ее услуг составлены так, что получить возмещение в случае порчи или кражи вложения невозможно. Однако несомненным плюсом скоростной почты является практически полное отсутствие хищений из почтовых отправлений.

---

[2] Сайт «РБК: Личные финансы». – http://If.rbc.ru/news/other/2011/03/30/178458.shtml.

Кроме почтовых служб, на российском рынке активно действует большое число крупных частных транспортных компаний, которые имеют круглосуточные бесплатные горячие линии для приема заказов, склады в большинстве региональных центров и самое главное – доступные тарифы. Достаточно сказать, что стоимость доставки крупногабаритных грузов в разы дешевле, чем пересылка аналогичных объемов через *ФГУП «Почта России»* (например, *Транспортная компания «Кит»*; *http://tk-kit.ru*).

Единственным крупным недостатком отправки товаров при помощи частных транспортных компаний является то, что товар доставляется на склад компании, откуда получатель забирает его самовывозом. Аналогичным образом осуществляется доставка груза до склада транспортной компании отправителем. Грузоотправитель самостоятельно должен подготовить все сопроводительные документы и сдать груз в упаковке, обеспечивающей его сохранность при перевозке.

В целом следует отметить, что услуги *ФГУП «Почта России»* оправданы, когда речь идет о недорогих мелкогабаритных товарах. Дорогие, хрупкие, ценные и габаритные товары целесообразно отправлять скоростной почтой (EMS обычно дешевле). Крупногабаритные товары имеет смысл отправлять при помощи частных транспортных компаний. Сегодня нет больших проблем с организацией доставки даже крупногабаритных грузов не только по России, но и из-за рубежа (Европа, Китай, Япония, США).

**Правовые основы дропшиппинга**. Поскольку дропшиппинг юридически мало отличается от несанкционированной торговли бабушек на блошиных рынках, то основным инструментом правового регулирования здесь выступает *Таможенный кодекс РФ*. Государство физически неспособно отследить каждого из миллионов интернет-пользователей. Да и затраты на обеспечение собираемости налоговых сборов вряд ли компенсировались бы этими сборами.

Что же касается таможенного регулирования, то ситуация для дропшипперов существенно улучшилась с введением с 6 июля 2010 г. нового *Таможенного кодекса РФ*. Согласно этому нормативному акту в течение одного календарного месяца в адрес одного получателя в международных почтовых отправлениях можно беспошлинно пересылать товары, совокупная таможенная стоимость которых не превышает 1.000 евро, а общий вес – 31 кг. В случае превышения указанных норм получатель оплачивает таможенные сборы по единой ставке 30% от таможенной стоимости товара, но не менее 4 евро за 1 кг веса в части превышения стоимостной нормы в 1.000 евро и (или) весовой нормы в 31 кг.

Это означает, что если в семье конечного получателя три члена, включая малолетнего ребенка, то указанные показатели можно смело умножать на три. Единственное ограничение – получаемые товары должны предназначаться исключительно для личного пользования. Иначе говоря, переслать 10 сотовых телефонов в одном отправлении беспошлинно будет сложно. Хотя и здесь есть лазейка: таможенная пошлина не будет взиматься, если 10 человек лично предъявят для таможенного оформления свои паспортные данные и договор-поручение с получателем посылки о приобретении на их средства 10 сотовых телефонов.

Тем более что дропшипперы вообще не выступают в качестве получателей товаров. Товары каждый раз приходят на адрес конечного получателя, и дропшиппер юридически как участник сделки в таможенных документах вообще не фигурирует. 10 сотовых телефонов здесь сразу идут на 10 не связанных между собой адресов. Поэтому дропшиппинг может вполне легально использоваться при продаже любых товаров, кроме запрещенных *Таможенным кодексом РФ* к пересылке в международных почтовых отправлениях.

В остальном дропшиппинг – абсолютно неконтролируемая сфера предпринимательства со стороны государства в условиях современной глобальной экономики. Специфика дропшиппинга не позволяет государственным фискальным органам отслеживать такого рода сделки, но зато открывает большие перспективы для решения, к примеру, проблем обеспечения жизнедеятельности отдаленных сельских районов и депрессивных территорий. С ростом доступности сети Интернет и наведением порядка в *ФГУП «Почта России»* этот ассортимент станет доступным не только в крупных городах, но и в сельской местности.

**Социально-экономические аспекты дропшиппинга**. Современный глобальный маркетинг предоставляет возможность обслуживания клиентов независимо от расстояния и наличия развитой торговой инфраструктуры.

Покупателям дропшиппинг дает доступ к широчайшему ассортименту товаров, который никогда не сможет обеспечить ни один, даже самым крупный, гипермаркет. Он повышает степень доступности товаров и обеспечивает равный доступ на рынки для производителей, минимизируя цены и повышая тем самым жизненный уровень покупателей.

Посредникам дропшиппинг дает возможность дополнительного заработка, что немаловажно в условиях экономического кризиса, особенно на депрессивных территориях. Вся прелесть заключается в том, что производитель может находиться в Китае, покупатель – в Австралии, а посредник снимет свою прибыль в российском заштатном городишке.

Торговым площадкам дропшиппинг дает дополнительные доходы, а также возможность в десятки раз увеличить количество участников и ассортимент предлагаемых товаров. Это может стать дополнительным стимулом для развития интернет-торговли в России. Ненормально, когда крупнейшая в СНГ российская торговая площадка зарегистрирована в Голландии, а обслуживается польскими программистами.

Производителям дропшиппинг дает колоссальные возможности для выхода на мировые рынки. Теперь никто не может сказать, что зарубежные рынки недоступны для отечественных производителей или требуют больших затрат на освоение. С дропшиппингом расстояния вообще утрачивают всякое значение, так как цена на почтовую отправку бандероли в Казахстан или в Новую Зеландию одинакова. Причем, это в равной мере относится к продаже, как наукоемкой продукции, так и изделий народных промыслов.

Государству дропшиппинг позволяет снизить остроту социальных проблем в условиях экономического кризиса за счет самозанятости населения и увеличения доступности потребительских товаров.

Разумеется, тут есть и негативные аспекты, связанные с некоторым падением в результате развития дропшиппинга бюджетных доходов и деградацией традиционной торговли. Однако дропшиппинг – объективная закономерность экономического развития в условиях бурного развития информационных технологий, которая уже сейчас коренным образом меняет всю инфраструктуру потребительского рынка. Именно сегодня, когда этот процесс только набирает обороты, существует острая необходимость на государственном уровне предпринять усилия по всецелой поддержке и пропаганде этой новой, по сути, формы продвижения товаров на потребительском рынке.

В конце концов, ведущие страны мира не имеют в большинстве своем запасов природных ресурсов, сопоставимых с российскими, а уровень жизни населения там существенно выше. Основной капитал отечественной экономики – предпринимательская активность населения – используется пока намного меньше, чем это могло бы быть. Тогда как дропшиппинг открывает огромные возможности для пред-принимательской деятельности на качественно более высоком уровне социальной организации с абсолютно равными возможностями для всех участников товародвижения.

**Библиографический список:**

1. Холленсен С. Глобальный маркетинг. – М.: Новое знание, 2004.